\documentclass[10pt, a4paper]{article}
\usepackage{amsmath,amssymb}

\usepackage{authblk}
\usepackage{enumitem}

\newtheorem{theorem}{Theorem}

\newcommand{\benumerate}{\begin{enumerate}}
\newcommand{\eenumerate}{\end{enumerate}}

\newcommand{\bitemize}{\begin{itemize}}

\newcommand{\eitemize}{\end{itemize}}

\begin{document}

\vspace{-2cm}
\title{Higher-order reductions of the Mikhalev system}
\author{E.V. Ferapontov$^{1,2}$,
V.S. Novikov$^{1}$,
I. Roustemoglou$^{1}$
}
     \date{}
     \maketitle
     \vspace{-10mm}
\begin{center}
$^1$Department of Mathematical Sciences, Loughborough University \\
Loughborough, Leicestershire LE11 3TU, United Kingdom \\
\vspace{2mm}
$^2$ Institute of Mathematics, Ufa Federal Research Centre\\
Russian Academy of Sciences\\
112 Chernyshevsky Street,  Ufa 450077, Russia\\
    [2ex]
e-mails: \\[1ex]
\texttt{E.V.Ferapontov@lboro.ac.uk}\\
\texttt{V.Novikov@lboro.ac.uk}\\
\texttt{I.Roustemoglou@lboro.ac.uk}

\end{center}

\medskip

\begin{abstract}

\end{abstract}

We consider the 3D
Mikhalev system, 
$$
u_t=w_x, \quad u_y= w_t-u w_x+w u_x, 
$$
which has first appeared in the context of KdV-type hierarchies. Under the reduction $w=f(u)$, one obtains a pair of commuting first-order equations,
$$
u_t=f'u_x, \quad u_y=(f'^2-uf'+f)u_x,
$$
which govern simple wave solutions of the Mikhalev system. In this paper we study {\it higher-order} reductions of the form
$$
w=f(u)+\epsilon a(u)u_x+\epsilon^2[b_1(u)u_{xx}+b_2(u)u_x^2]+...,
$$
which turn the Mikhalev system into a pair of commuting higher-order equations. Here the terms at $\epsilon^n$  are assumed to be differential polynomials of degree $n$  in the $x$-derivatives of $u$. We will view $w$ as an  (infinite) formal series in the deformation parameter $\epsilon$. It turns out that for such a reduction to be non-trivial, the function $f(u)$ must be quadratic, $f(u)=\lambda u^2$, furthermore, the value of the parameter $\lambda$ (which has a natural interpretation as an eigenvalue of a certain second-order operator acting on an infinite jet space), is quantised. There are only two positive allowed eigenvalues, $\lambda=1$  and $\lambda=3/2$, as well as infinitely many negative rational eigenvalues.

Two-component  reductions of the Mikhalev system are also discussed. We emphasise that the existence of higher-order reductions of this kind is a reflection of {\it linear degeneracy} of the Mikhalev system, in particular, such reductions do not exist for most of the known 3D dispersionless integrable systems such as the dispersionless KP and Toda equations. 

\bigskip

\bigskip


\medskip


\newpage

\tableofcontents

\section{Introduction}
\label{sec:intro}

We consider the 3D
Mikhalev system, 
\begin{equation}
\label{pav}
u_t=w_x, \qquad u_y= w_t-u w_x+w u_x, 
\end{equation}
which was apparently first introduced by V. Mikhalev  \cite{Mikhalev} in the context of Hamiltonian formalism of KdV type hierarchies, and subsequently investigated in the framework of multi-dimensional dispersionless integrability, see e.g.  \cite{Dun, Pavlov1, FM, FKK} and references therein. We will be interested in the classification of $(1+1)$-dimensional reductions of (\ref{pav})  obtained by setting 
\begin{equation}
w=f(u)+\epsilon a(u)u_x+\epsilon^2[b_1(u)u_{xx}+b_2(u)u_x^2]+...
\label{w}
\end{equation}
where the terms at $\epsilon^n$  are differential polynomials of degree $n$  in the $x$-derivatives of $u$  (whose coefficients are some functions of $u$). We will view (\ref{w}) as an  infinite formal series in the deformation parameter $\epsilon$. Reduction  (\ref{w}) will be called nontrivial if at least one of the $\epsilon$-dependent terms is nonzero. 
The substitution of (\ref{w}) into (\ref{pav}) implies
\begin{equation}
u_t=w^*u_x, \qquad u_y=((w^*)^2-uw^*+w)u_x,
\label{*}
\end{equation}
 where $w^*=w_u+w_{u_x}\frac{d}{dx}+w_{u_{xx}}\frac{d^2}{dx^2}+\dots$ is the operator of formal linearisation. Thus, we obtain  two equations  for the scalar variable $u$.  Requiring their  compatibility,  $u_{yt}=u_{ty}$ (which should be satisfied at all orders of the deformation parameter $\epsilon$), one obtains strong constraints for the coefficients of expansion (\ref{w}). There are only two cases of terminating expansions, leading to the well-known integrable hierarchies.
 
 \medskip
 
 \noindent{\bf Example 1. } The ansatz
 $ w=u^2+\epsilon u_x$ reduces (\ref{pav}) to the first two flows of the Burgers' hierarchy,
 $$
 u_t=2uu_x+\epsilon u_{xx}, \qquad u_y=3u^2u_x+3\epsilon(u_x^2+uu_{xx})+\epsilon^2 u_{xxx}.
 $$

\medskip
 
 \noindent{\bf Example 2. } The ansatz
 $ w=\frac{3}{2}u^2+\epsilon^2 u_{xx}$ reduces (\ref{pav}) to the first two flows of the KdV hierarchy,
 $$
 u_t=3uu_x+\epsilon^2 u_{xxx}, \qquad u_y=\frac{15}{2}u^2u_x+5\epsilon^2(uu_{xxx}+2u_xu_{xx})+\epsilon^4 u_{xxxxx}.
 $$
In the language of the `universal hierarchy of hydrodynamic type' and `integrable hydrodynamic chains', terminating reductions of this kind were studied in \cite{MAS1, MAS2, MAS3, Pavlov1}, see also \cite{Pavlov21} for some explicit  finite-gap solutions of the Mikhalev system. Our first main result is as follows (see Section \ref{sec:T1} for the proof):

\begin{theorem} \label{T1} For  a non-trivial reduction (\ref{w}) of system (\ref{pav}), the function $f(u)$ must be quadratic. 

\end{theorem}

\medskip

Thus, one can set $f(u)=\alpha+\beta u+\lambda u^2$. Here the parameters $\alpha, \beta$ are not essential and can be eliminated using the following symmetries of system (\ref{pav}):
\medskip

\noindent (a): $\tilde x=x+\alpha y, \ \tilde y=y, \ \tilde t=t, \ \tilde u=u, \ \tilde w=w-\alpha$;

\noindent (b): $\tilde x=x+\beta^2 y+\beta t, \ \tilde y=y, \ \tilde t=t+2\beta y, \ \tilde u=u, \ \tilde w=w-\beta u$.
\medskip

 Therefore, without any loss of generality one can assume $f(u)=\lambda u^2$. Further analysis of Section \ref{sec:lambda} shows that the allowed values of the parameter $\lambda$ are `quantised' and cannot be arbitrary. There are only two positive eigenvalues, $\lambda=1$ are $\lambda=\frac{3}{2}$. 
Computer experiments suggest that there exist infinitely many series of negative rational eigenvalues $\lambda$, in particular,
$$
\begin{array}{c}
\lambda=\frac{2}{2(2-n)} \ (n\geq 3), \qquad \lambda=\frac{3}{2(3-n)} \ (n\geq 4), \qquad \lambda=\frac{4}{2(4-n)} \ (n\geq 5), \\
\ \\
\lambda=\frac{5}{2(5-n)} \ (n\geq 6), \qquad \lambda=\frac{6}{2(6-n)} \ (n\geq 8), \qquad \lambda=\frac{7}{2(7-n)} \ (n\geq 10), 
\end{array}
$$
etc, although we were not able to find a  general formula giving all of them.

In Section \ref{sec:red} we study in more detail  deformations (\ref{w}) corresponding to the two positive values of $\lambda$. Thus, the general expansion (\ref{w}) for $\lambda=1$ (dissipative case) has the form 
$$
w=u^2+\epsilon au_x+\frac{\epsilon^2}{2} (a^2)_{xx}+\frac{\epsilon^3}{2}\Big[(a^2a')_{xx}-(aa'a''+\frac{1}{6}a^2a''')u_x^2\Big]_x+\dots,
$$
where $a(u)$ is an arbitrary function. For $a=1$ it reproduces the Burgers' reduction from  Example 1. For $a=u$ one obtains another well-known case, the so-called viscous Camassa-Holm reduction, see Section \ref{sec:dissip} for more details. Similarly, the general expansion (\ref{w}) for $\lambda=\frac{3}{2}$ (dispersive case) has the form
  \begin{align*}
 w&=\frac{3}{2}u^2+\epsilon^2(au_{xx}+\frac{5}{4}a'u_x^2)+ \frac{\epsilon^4}{128}\left( 64 a a' u_{xxxx}  +32  (9 {a'}^2+7 a a'')u_x u_{xxx} +\right .\\
&~~~~~~\left .+16  (14 {a'}^2+9 a a'')u_{xx}^2 +96 \left(13 a' a''+3 a a'''\right) u_x^2 u_{xx} +5 (52 a''' a'+44 {a''}^2+9 a a^{(4)})u_x^4 \right)+\dots,
 \end{align*}
where $a(u)$ is an arbitrary function. For $a=1$ it reproduces the KdV reduction from  Example 2.  For $a=u$ one obtains the  Camassa-Holm reduction, see Section \ref{sec:disper} for more details. 

In Section \ref{sec:two-comp} we study higher-order (dissipative and dispersive) deformations of two-component hydrodynamic reductions of the Mikhalev system (\ref{pav}). 

Finally, in Section \ref{sec:dKP} we prove non-existence of reductions of type (\ref{w}) for the dispersionless KP equation and justify a conjecture that only {\it linearly degenerate} dispersionless integrable systems in 3D possess higher-order reductions of this kind.

\section{Proof of Theorem \ref{T1}}
\label{sec:T1}

Here we prove our first main result.

\medskip

\noindent{\bf Theorem \ref{T1}} {\it For  a non-trivial reduction (\ref{w}) of system (\ref{pav}), the function $f(u)$ must be quadratic. }

\medskip

\centerline{\bf Proof:}

\medskip

If $f$ is not quadratic, direct calculations show that terms at $\epsilon, \epsilon^2, \epsilon^3, \dots$ in the expansion (\ref{w}) must vanish identically. Suppose that the first nonzero term occurs at $\epsilon^n$,
$$
w=f(u)+\epsilon^nw_n+...,
$$
where $w_n$ is a homogeneous differential polynomial of differential degree $n$ in the $x$-derivatives of $u$. Equations (\ref{pav}) take the form
\begin{equation}
\begin{array}{c}
u_t=f'u_x+\epsilon^n w_{n, x}+\dots, \\
\ \\
u_y=(f'^2-uf'+f)u_x+\epsilon^n[w_nu_x+(f'-u)w_{n, x}+w_{n, t}]+\dots.
\end{array}
\label{f}
\end{equation} 
Calculating the compatibility condition of (\ref{f}), $u_{yt}=u_{ty}$, at $\epsilon^n$ one obtains that $w_n$ must 
satisfy the linear equation
\begin{equation}\label{wn}
w_{n, tt}-w_{n, xy}-uw_{n, xt}+fw_{n, xx}+(1-f'')u_x(w_{n, t}-f'w_{n,x})=0,
\end{equation}
where one has to use dispersionless limits of equations (\ref{f}),   $u_t=f'u_x$ and $u_y=(f'^2-uf'+f)u_x$, to calculate $t$- and $y$-derivatives of $w_n$.  Introducing the corresponding commuting operators of total differentiation, 
$$
\begin{array}{c}
X=u_x\frac{\partial}{\partial u}+u_{xx}\frac{\partial}{\partial u_x}+\dots, \\
\ \\
T=f'u_x\frac{\partial}{\partial u}+(f'u_x)_x\frac{\partial}{\partial u_x}+\dots,\\
\ \\
Y=(f'^2-uf'+f)u_x\frac{\partial}{\partial u}+((f'^2-uf'+f)u_x)_x\frac{\partial}{\partial u_x}+\dots, \\
\end{array}
$$
one can rewrite equation (\ref{wn}) in the form $Lw_n=0$ where $L$ is a second-order operator,
$$
L=T^2-YX-uTX+fX^2+(1-f'')u_x(T-f'X);
$$
note that $L$ is independent of $n$. It is convenient to introduce the operators 
$$
\tilde T=T-f'X, \quad \tilde Y=Y-(f'^2-uf'+f)X, \quad S=(2f'-u)\tilde T-\tilde Y,
$$
note that $\tilde T$ and $\tilde Y$ do not involve differentiation $\partial_u$, furthermore, 
$S$ does not involve differentiations $\partial_u$ and $\partial_{u_x}$. In this notation,  $L$ takes the form
\begin{equation}\label{L}
L=\tilde T^2+SX+(1-f'')u_x\tilde T;
\end{equation}
We also note that $L$ does not explicitly depend on $f$ and $f'$. Let us introduce the standard notation for the jet variables: $(u, u_x, u_{xx}, \dots)=(u, u_1, u_2, \dots)$. Each differential monomial has two naturally defined degrees: 
differential degree $n$ and algebraic degree $k$.  Thus, the differential monomial $u_{\sigma_1} \dots  u_{\sigma_k}$ has differential degree $n=\sigma_1+\dots+\sigma_k$  and algebraic degree k. Let $P_k(n)$ be the space of differential polynomials of differential degree $n$ and algebraic degree $k$, spanned by the corresponding differential monomials. Then ${\rm dim}\, P_k(n)=p_k(n)$ where $p_k(n)$, the number of differential monomials of differential degree $n$ and algebraic degree $k$, equals the number of partitions of $n$ into $k$ parts. We will agree that the variable $u$ has both differential and algebraic degrees equal to zero.
We will also need the following properties of operators  $\tilde T, S$ and $X$ entering formula (\ref{L}). First of all, each of these operators increases by $1$ the differential degree of any differential monomial. It will be convenient to represent $X$ in the form $X=X_0+u_1\frac{\partial}{\partial u}$ where $X_0=u_{2}\frac{\partial}{\partial u_1}+u_{3}\frac{\partial}{\partial u_2}+\dots$ preserves algebraic degree of any differential  monomial, while $u_1\frac{\partial}{\partial u}$ increases it by $1$. Similarly, we have
$\tilde T=\tilde T_0+\dots$  where
$$
\tilde T_0=f''\left(u_1^2 \frac{\partial}{\partial u_1}+3u_1u_2\frac{\partial}{\partial u_2}+(4u_1u_3+3u_2^2)\frac{\partial}{\partial u_3}+\dots\right);
$$
note that $\tilde T_0$ increases by $1$ both algebraic and differential degrees of any differential monomial. Here dots in the formula  $\tilde T=\tilde T_0+\dots$ denote an operator which increases algebraic degree by more than $1$. Finally, $ S=S_0+\dots$ where
$$
S_0=f''(1-2f'')\left(u_1^3 \frac{\partial}{\partial u_2}+6u_1^2u_2\frac{\partial}{\partial u_3}+\dots\right);
$$
note that $S_0$ increases by $2$ the algebraic degree, and by $1$ the differential degree of any differential monomial. Here dots in the formula  $S=S_0+\dots$ denote an operator which increases algebraic degree by more than $2$. Thus, one can represent the second-order operator $L$ given by (\ref{L}) in the form
$
L=L_0+\dots
$
where
\begin{equation}\label{L0}
L_0=\tilde T_0^2+S_0X_0+(1-f'')u_1\tilde T_0;
\end{equation}
note that $L_0$ increases by $2$ both algebraic and differential degrees of any differential monomial. Here dots in the formula  $L=L_0+\dots$ denote a second-order operator which increases algebraic degree by more than $2$. Finally, let us note that $L_0$ depends on $f$ only via the second derivative $f''$.

Our goal is to show that, if $f''\ne const$,  the kernel of $L$ is trivial: $Lw_n=0\implies w_n=0$. To prove this we represent $w_n$, which is a differential polynomial of differential degree $n$, in the form 
\begin{equation}\label{exp}
w_n=w_n^{(1)}+w_n^{(2)}+w_n^{(3)}+\dots,
\end{equation}
 where $w_n^{(i)}$ are differential polynomials of differential degree $n$ and algebraic degree $i$: 
 $$
 w_n^{(1)}={\rm span} \langle u_n\rangle, \quad w_n^{(2)}={\rm span} \langle u_{n-s}u_s\rangle,
 $$
 etc. Here  coefficients in the span are allowed to be arbitrary functions of $u$.  Suppose $w_n^{(k)}$ is the first nonzero term in  expansion (\ref{exp}). Then the term of lowest algebraic degree in the expression $Lw_n$ will be $L_0w_n^{(k)}$. As $Lw_n=0$ implies $L_0w_n^{(k)}=0$, it is sufficient to show that 
 $L_0w_n^{(k)}=0\implies w_n^{(k)}=0$. 
 
 The condition $L_0w_n^{(k)}=0$ gives a linear homogeneous system for the coefficients of differential polynomial $w_n^{(k)}$ (recall that $L_0$ does not involve differentiation by $u$); the matrix of this system depends on $f''$ which can be treated as an arbitrary `parameter'. The requirement of non-trivial solvability of this system leads to two cases: either there is an algebraic constraint for $f''$ (in which case we are done as $f''$ will be constant), or the system is solvable for every value of $f''$. The latter case leads to a contradiction: if the system is non-trivially solvable for every value of $f''$, let us set $f''=\frac{1}{2}$. In this case the operator $S_0$ vanishes identically, and the condition  $L_0w_n^{(k)}=0$ simplifies to $(\tilde T_0^2+\frac{1}{2}u_1\tilde T_0)w_n^{(k)}=0$. Note however that the lexicographically top monomial in the expression $(\tilde T_0^2+\frac{1}{2}u_1\tilde T_0)w_n^{(k)}$ equals the lexicographically top monomial of $w_n^{(k)}$ multiplied by $cu_1^2$ where $c$ is some positive constant (as all coefficients of the operator 
 $\tilde T_0^2+\frac{1}{2}u_1\tilde T_0$ are positive). Thus, in this case the vanishing of $L_0w_n^{(k)}$ implies the vanishing of $w_n^{(k)}$.
 This contradiction finishes the proof. 
 
 $\hfill \square$

\section{Eigenvalues $\lambda$}
\label{sec:lambda}

Upon setting $f(u)=\lambda u^2$, it follows from the proof of Theorem \ref{T1} that the allowed values of the parameter $\lambda$ are exactly those for which the operator $L_0$ given by (\ref{L0}) possesses a nontrivial kernel. The corresponding $\lambda$'s can be interpreted as the eigenvalues of $L_0$. As the operator $L_0$ increases by $2$ both the differential and algebraic degrees of every differential monomial, one needs to study a finite-dimensional restriction of the operator $L_0$ to the space $P_k(n)$:
\begin{equation}\label{pkn}
L_0:\, P_k(n)\to P_{k+2}(n+2).
\end{equation}
Our first observation is that a sufficient condition for the kernel of $L_0$ to be non-trivial is the equality of dimensions:
\begin{equation}\label{part}
p_k(n)=p_{k+2}(n+2);
\end{equation}
in this case, elements of the kernel correspond to the eigenvalues $\lambda$ that satisfy the equation $\det L_0=0$ (these eigenvalues depend on both $n$ and $k$). Note that there is an obvious general inequality $p_k(n)\leq p_{k+2}(n+2)$, indeed, every differential monomial from $P_k(n)$ produces a differential monomial from $P_{k+2}(n+2)$ via multiplication by $u_1^2$. One can easily show that the equality (\ref{part}) holds if and only if $k\geq n/2$. This follows from the well-known identity for partitions, $p_k(n)=p_k(n-k)+p_{k-1}(n-1)$. Applying this identity twice gives
$$
p_{k+2}(n+2)=p_{k+2}(n-k)+p_{k+1}(n+1)=p_{k+2}(n-k)+p_{k+1}(n-k)+p_{k}(n),
$$
where the first two terms on the right-hand side vanish if and only if $k\geq n/2$ (recall that $p_k(n)=0$ if $k>n$). 


\medskip

Our results are summarised in Table 1 below: the left column lists  the first few allowed eigenvalues  $\lambda$; the right column contains the first two terms in the expansion of $w(u)$. 

\newpage
\begin{center}
\centerline{\footnotesize{Table 1: Deformations starting with $\epsilon^k, \, k \leq 6$}}
 \begin{tabular}{ | l | l | l | p{1.1cm} |} \hline
 {\footnotesize  $\lambda$}  & {\footnotesize Expression for $w(u)$}  \\
 \hline 
{\footnotesize $1$}&\footnotesize{$w(u)=u^2+\epsilon au_x+\dots$}  \\
\hline

{\footnotesize $ \frac{3}{2}$}&\footnotesize{$w(u)=\frac{3}{2}u^2+\epsilon^2(au_{xx}+\frac{5}{4}a'u_x^2) +\dots$}   \\
\hline

{\footnotesize $-1$}&\footnotesize{$w(u)=-u^2+\epsilon^3(au_{x}^3) +\dots$}   \\
\hline

{\footnotesize $-\frac{1}{3}$}&\footnotesize{$w(u)=-\frac{1}{3}u^2+\epsilon^4(au_{xxx}u_x-3au_{xx}^2) +\dots$}   \\
\hline

{\footnotesize $-\frac{3}{2}$}&\footnotesize{$w(u)=-\frac{3}{2}u^2+\epsilon^4(au_{xx}u_x^2+\frac{7}{16}a'u_x^4) +\dots$}   \\
\hline

{\footnotesize $-\frac{1}{2}$}&\footnotesize{$w(u)=-\frac{1}{2}u^2+\epsilon^4(au_x^4) +\dots$}   \\
\hline

{\footnotesize $-{2}$}&\footnotesize{$w(u)=-{2}u^2+\epsilon^5(au_{xxx}u_x^2+\frac{5}{3}au_{xx}^2u_x+\frac{14}{5}a'u_{xx}u_x^3+\frac{63}{125}a''u_x^5) +\dots$}   \\
\hline

{\footnotesize $-\frac{1}{4}$}&\footnotesize{$w(u)=-\frac{1}{4}u^2+\epsilon^5(au_{xxx}u_x^2-{3}au_{xx}^2u_x) +\dots$}   \\
\hline

{\footnotesize $-\frac{3}{4}$}&\footnotesize{$w(u)=-\frac{3}{4}u^2+\epsilon^5(au_{xx}u_x^3+\frac{8}{25}a'u_x^5) +\dots$}   \\
\hline

{\footnotesize $-\frac{1}{3}$}&\footnotesize{$w(u)=-\frac{1}{3}u^2+\epsilon^5(au_x^5) +\dots$}   \\
\hline

{\footnotesize $-\frac{3}{8}$}&\footnotesize{$w(u)=-\frac{3}{8}u^2+\epsilon^6(au_{xxxx}u_x^2-\frac{51}{10}au_{xxx}u_{xx}u_x+\frac{3}{10}au_{xx}^3+a'u_{xxx}u_x^3-3a'u_{xx}^2u_x^2) +\dots$}   \\
\hline

{\footnotesize $-\frac{1}{5}$}&\footnotesize{$w(u)=-\frac{1}{5}u^2+\epsilon^6(au_{xxxx}u_x^2-{10}au_{xxx}u_{xx}u_x+15au_{xx}^3+bu_{xxx}u_x^3-3bu_{xx}^2u_x^2) +\dots$}   \\
\hline

{\footnotesize $-\frac{5}{2}$}&\footnotesize{$w(u)=-\frac{5}{2}u^2+\epsilon^6(au_{xxxx}u_x^2+\frac{51}{11}au_{xxx}u_{xx}u_x+\frac{454}{363}au_{xx}^3+\frac{23}{6}a'u_{xxx}u_x^3+$} \\
& {\footnotesize $\qquad\qquad\qquad\qquad\qquad\qquad\qquad\qquad +\frac{1219}{132}a'u_{xx}^2u_x^2+\frac{391}{72}a''u_{xx}u_x^4+\frac{4301}{7776}a'''u_x^6) +\dots$}   \\
\hline

{\footnotesize $-1$}&\footnotesize{$w(u)=-u^2+\epsilon^6(au_{xxx}u_x^3+\frac{13}{5}u_{xx}^2u_x^2+\frac{8}{3}a'u_{xx}u_x^4+\frac{10}{27}a''u_x^6) +\dots$}   \\
\hline

{\footnotesize $-\frac{1}{2}$}&\footnotesize{$w(u)=-\frac{1}{2}u^2+\epsilon^6(au_{xx}u_x^4+\frac{1}{4}a'u_x^6) +\dots$}   \\
\hline

{\footnotesize $-\frac{1}{4}$}&\footnotesize{$w(u)=-\frac{1}{2}u^2+\epsilon^6(au_x^6) +\dots$}   \\
\hline
\end{tabular}
\end{center}

\noindent Here $a, b$ are arbitrary function of $u$ and dots denote higher-order terms in $\epsilon$. Note that the value $\lambda=-1$ appears at the orders $\epsilon^3$ and $\epsilon^6$, the value $\lambda=-\frac{1}{3}$ appears at the orders $\epsilon^4$ and $\epsilon^5$, etc. This means that the spectrum of the operator $L_0$ is not simple. Furthermore,  for $\lambda=-\frac{1}{5
}$ two arbitrary functions, $a$ and $b$, appear in the expression for $w$ at the order $\epsilon^6$. This means that the operator $L_0$ does not have simple spectrum even when restricted to the subspace  of differential polynomials of a fixed degree $n$.

\medskip

Based on Table 1, we have found the following  deformations that work for arbitrary value of $n$.

\medskip

\begin{center}
 \centerline{\footnotesize{Table 2: Examples of deformations of arbitrary order $n$}}
 \medskip
 \begin{tabular}{ | l | l | l | p{1.1cm} |} \hline
 {\footnotesize  $\lambda$}  & {\footnotesize Expression for $w(u)$}  \\
 \hline

{\footnotesize $\frac{1}{2-n}$}&\footnotesize{$w(u)=\frac{1}{2-n}u^2+\epsilon^n (au_x^n)+\dots$}, ~~ $n\ne 2$  \\
\hline

{\footnotesize $ \frac{3}{2(3-n)}$}&\footnotesize{$w(u)=\frac{3}{2(3-n)}u^2+\epsilon^n(au_{xx}u_x^{n-2}+\frac{n+3}{n^2}a'u_x^n) +\dots$}, ~~ $n\ne1, 3$   \\
\hline

{\footnotesize $\frac{1}{1-n}$}&\footnotesize{$w(u)=\frac{1}{1-n}u^2+\epsilon^n(au_{xxx}u_{x}^{n-3}-3au_{xx}^2u_x^{n-4}) +\dots$}, ~~ $n\geq 4$   \\
\hline

{\footnotesize $\frac{2}{4-n}$}&\footnotesize{$w(u)=\frac{2}{4-n}u^2+\epsilon^n(au_{xxx}u_x^{n-3}+\frac{n^2-10}{n+4}au_{xx}^2u_x^{n-4}+2\frac{n+2}{n}a'u_{xx}u_x^{n-2}+\frac{(n+2)(n+4)}{n^3}a''u_x^n) +\dots$}, ~~ $n\geq 5$   \\
\hline

\end{tabular}

\end{center}

Computer experiments (up to  $n=12$ and all $k\leq n$) suggest that there exist infinitely many series of negative rational eigenvalues $\lambda$, in particular,
$$
\begin{array}{c}
\lambda=\frac{2}{2(2-n)} \ (n\geq 3), \qquad \lambda=\frac{3}{2(3-n)} \ (n\geq 4), \qquad \lambda=\frac{4}{2(4-n)} \ (n\geq 5), \\
\ \\
\lambda=\frac{5}{2(5-n)} \ (n\geq 6), \qquad \lambda=\frac{6}{2(6-n)} \ (n\geq 8), \qquad \lambda=\frac{7}{2(7-n)} \ (n\geq 10), 
\end{array}
$$
etc, although we were not able to find a  general formula that would give  all of them.

\section{Higher-order deformations of one-component reductions}
\label{sec:red}

Here we study in more detail higher-order reductions corresponding to the two positive values of $\lambda$. 

\subsection{One-component dissipative reductions $(\lambda=1)$}
\label{sec:dissip}

In this case   expansion (\ref{w}) takes the form
$$
w=u^2+\epsilon au_x+\frac{\epsilon^2}{2} (a^2)_{xx}+\frac{\epsilon^3}{2}\Big[(a^2a')_{xx}-(aa'a''+\frac{1}{6}a^2a''')u_x^2\Big]_x+\dots,
$$
here $a(u)$ is an arbitrary function, and the coefficients of all subsequent terms are certain explicit expressions in  $a(u)$ and its derivatives (in both the dissipative and dispersive cases, we have performed calculations of $w$ and verified commutativity of the corresponding flows (\ref{*}) up to the order $\epsilon^8$).  The corresponding equation $u_t=w_x$ belongs to the class of `viscous conservation laws' studied in \cite{ALM1, ALM2}. The coefficient $a(u)$ is referred to as the `viscous central invariant' (note that our normalisation is different from that of \cite{ALM1, ALM2}). This expansion does not terminate unless $a(u)=const$, which corresponds to the Burgers' hierarchy (Example 1 of Section \ref{sec:intro}).

Another familiar case is $a(u)=u$ which gives
$$
w=u^2+\frac{\epsilon}{2}(u^2)_x+\frac{\epsilon^2}{2}(u^2)_{xx}+\frac{\epsilon^3}{2}(u^2)_{xxx}+\dots.
$$
Although this expansion does not terminate, it can be represented in compact form as
\begin{equation}\label{w2}
w=u^2+\frac{\epsilon}{2}\frac{ \partial_x}{1-\epsilon \partial_x}(u^2). 
\end{equation}
Introducing $m=u-\epsilon u_x$, one can rewrite the corresponding equations (\ref{pav}) in the form
\begin{equation}\label{CH2}
m_t=(mu)_x, \quad m_y=(mw)_x.
\end{equation}
Note that the first equation (\ref{CH2}) can be written in the following second-order non-evolutionary form,
$$
u_t-\epsilon u_{xt}=2uu_x-\epsilon(uu_{xx}+u_x^2).
$$
This equation was discussed in \cite{ALM2} as a viscous analogue of the Camassa-Holm equation.
It has first appeared in \cite{OR}, see also \cite{ F}.
To obtain non-evolutionary form of the second equation (\ref{CH2}), one can use the relation $w-\epsilon w_x=um$, which is equivalent to (\ref{w2}), to eliminate $w$. This results in the following third-order non-evolutionary form:
$$
\epsilon m_{xy}-\frac{\epsilon m_{xx}+m_x}{\epsilon m_x+m}(\epsilon m_y+um^2)=m_y-3umm_x-m^2u_x.
$$
Commuting flows (\ref{CH2}) belong to the hierarchy of the second-order linearisable PDE
$$
u_{\tau}=\left(\frac{1}{u-\epsilon u_x}\right)_x, \quad {\rm equivalently}, \quad m_{\tau}=(\partial_x-\epsilon \partial_x^2)\left(\frac{1}{m}\right),
$$
which has appeared in \cite{Calogero}, see also \cite{Novikov}, eq. (17).

Unfortunately, already the case $a(u)=u^2$ is not  representable in any reasonably compact form, furthermore, our calculations suggest that the corresponding commuting flows (\ref{*}) do not belong to the hierarchy of any finite-order integrable PDE.

\subsection{One-component dispersive reductions $(\lambda=3/2)$}
\label{sec:disper}

In this case  expansion (\ref{w}) takes the form
  \begin{align*}
 w&=\frac{3}{2}u^2+\epsilon^2(au_{xx}+\frac{5}{4}a'u_x^2)+ \frac{\epsilon^4}{128}\left( 64 a a' u_{xxxx}  +32  (9 {a'}^2+7 a a'')u_x u_{xxx} +\right .\\
&~~~~~~\left .+16  (14 {a'}^2+9 a a'')u_{xx}^2 +96 \left(13 a' a''+3 a a'''\right) u_x^2 u_{xx} +5 (52 a''' a'+44 {a''}^2+9 a a^{(4)})u_x^4 \right)+\dots,
 \end{align*}
 where $a(u)$ is an arbitrary function, and  all subsequent terms are certain explicit expressions in  $a(u)$ and its derivatives. 
  This expansion does not terminate unless $a(u)=const$, which corresponds to the KdV hierarchy (Example 2 of Section \ref{sec:intro}).
  
Another familiar case is  $a(u)=2u$ (\cite{Pavlov1}, Sect. 5) which gives
$$
w=\frac{3}{2}u^2+\epsilon^2\left(2u u_{xx}+\frac{5}{2}u_x^2\right)+\dots.
$$
Although this expansion does not terminate, it can be represented in compact form as
\begin{equation}\label{w3}
w=\left(1-\epsilon^2 \partial_x^2\right)^{-1}\left(-\epsilon^2u u_{xx}-\frac{\epsilon^2}{2}u_x^2+\frac{3}{2}u^2\right).
\end{equation}
Introducing $n=u-\epsilon^2 u_{xx}$, one can rewrite the corresponding equations (\ref{pav}) in the form
\begin{equation}\label{CH3}
n_t=(nu)_x+nu_x, \quad n_y=(nw)_x+nw_x.
\end{equation}
Note that the first equation (\ref{CH3}) can be written in the  Camassa-Holm non-evolutionary form,
$$
u_t-\epsilon^2u_{xxt}=3uu_x-\epsilon^2(u u_{xxx}+2u_xu_{xx}).
$$
To obtain non-evolutionary form of the second equation (\ref{CH3}), one can differentiate it by $x$ and use the relation $w-\epsilon^2 w_{xx}=un+\frac{1}{2}(u^2-\epsilon^2u_x^2)$, which is equivalent to (\ref{w3}), to eliminate $w$.
Commuting flows (\ref{CH3}) belong to the hierarchy of the third-order integrable PDE
$$
u_{\tau}=\left(\frac{1}{\sqrt{u-\epsilon^2 u_{xx}}}\right)_x, \quad {\rm equivalently}, \quad n_{\tau}=(\partial_x-\epsilon^2 \partial_x^3)\left(\frac{1}{\sqrt n}\right),
$$
which has appeared in \cite{CH}.


\section{Higher-order deformations of two-component reductions}
\label{sec:two-comp}

Multi-component hydrodynamic reductions of system (\ref{pav}) can be obtained by setting
$u=u(R^1, \dots, R^n)$, $w=w(R^1, \dots, R^n)$ where $R^i$ satisfy a pair of commuting $n$-component hyperbolic diagonal systems of hydrodynamic type,
\begin{equation}\label{R}
R^i_t=\eta^i(R) R^i_x, \quad R^i_y=\mu^i(R) R^i_x,
\end{equation}
whose characteristic speeds satisfy  the commutativity conditions  \cite{Tsa},
\begin{equation}\label{comm}
\frac{\partial_j\eta^i}{\eta^j-\eta^i}=\frac{\partial_j\mu^i}{\mu^j-\mu^i}, \quad i\ne j;
\end{equation}
here $\partial_i=\partial_{R^i}$, see \cite{FK} for the general theory of hydrodynamic reductions of multi-dimensional dispersionless integrable systems. The requirement that the so defined $u$ and $w$ solve system (\ref{pav}) imposes the following constraints for the unknown functions: $\partial_iw=\eta^i\partial_iu$ (no summation), as well as the dispersion relation $\mu^i=(\eta^i)^2-u\eta^i+w$. Upon substitution into commutativity conditions  (\ref{comm}) these constraints imply $\partial_j\eta^i=\partial_ju$, while the compatibility conditions $\partial_i\partial_jw=\partial_j\partial_iw$ imply $\partial_i\partial_ju=0$. These equations are straightforward to solve: with a suitable choice of Riemann invariants $R^i$, which are  defined up to reparametrisations $R^i\to \varphi^i(R^i)$, one has
\begin{equation}\label{hydro}
\begin{array}{c}
u=\sum R^k, \quad w=\frac{1}{2}\left(\sum R^k\right)^2+\sum f_k(R^k),\\
\ \\
\eta^i=\sum R^k+f_i'(R^i), \quad \mu^i=[f_i'(R^i)]^2+f_i'(R^i)\sum R^k+    \frac{1}{2}\left(\sum R^k\right)^2+\sum f_k(R^k);
\end{array}
\end{equation} 
here $f_i(R^i)$ are $n$ arbitrary functions of the indicated variables and all summations are from $1$ to $n$ (equivalent formulae were obtained in \cite{Pavlov}, Sect. 7).

Below we discuss higher-order deformations of  two-component  hydrodynamic reductions. Thus, we seek solutions of system (\ref{pav}) in the form
$$
u=R^1+R^2, \quad w=\frac{1}{2}(R^1+R^2)^2+f_1(R^1)+f_2(R^2),
$$
where $R^1, R^2$ satisfy equations of the form
\begin{equation}\label{Rdef}
R^i_t=\eta^i(R) R^i_x+\epsilon (\dots)+\epsilon^2(\dots)+\dots, \quad R^i_y=\mu^i(R) R^i_x+\epsilon (\dots)+\epsilon^2(\dots)+\dots,
\end{equation}
where dots at $\epsilon^i$ denote differential polynomials of degree $i+1$ whose coefficients can be arbitrary functions of $R^1, R^2$ ($u$ and $w$ can be assumed undeformed due to the invariance of the construction under the Miura group, see, e.g., \cite{DLZ}). The characteristic speeds $\eta^i, \ \mu^i$ are as follows:
$$
\begin{array}{c}
\eta^i=R^1+R^2+f_i'(R^i), \qquad \mu^i=[f_i'(R^i)]^2+f_i'(R^i)(R^1+R^2)+    \frac{1}{2}\left(R^1+R^2\right)^2+f_1(R^1)+f_2(R^2),
\end{array}
$$
here $i=1, 2$. The substitution of this ansatz into (\ref{pav}) implies that equations (\ref{Rdef}) can be written as
\begin{equation}\label{T}
R^1_t=\eta^1(R) R^1_x+P, \quad R^2_t=\eta^2(R) R^2_x-P
\end{equation}
and
\begin{equation}\label{Y}
R^1_y=\mu^1(R) R^1_x+f_1'(R^1)P+Q, \quad R^2_y=\mu^2(R) R^2_x-f_2'(R^2)P-Q,
\end{equation}
respectively. Here $P$ and $Q$ are expansions of the form
$$
P=\epsilon (\dots)+\epsilon^2(\dots)+\dots, \quad Q=\epsilon (\dots)+\epsilon^2(\dots)+\dots,
$$
where the terms at $\epsilon^i$ are differential polynomials of degree $i+1$ whose coefficients can be arbitrary functions of $R^1, R^2$.
We will see that the requirement of compatibility of systems (\ref{T}) and (\ref{Y}), $R^i_{ty}=R^i_{yt}$,  imposes strong constraints on the functions $f_i(R^i)$.

\subsection{Two-component dissipative reductions}
\label{sec:dissip2}

Here we assume that the expansions start with non-trivial $\epsilon$-terms,
$$
P=\epsilon [p_1R^1_{xx}+p_2R^2_{xx}+p_{3}(R^1_x)^2+p_{4}R^1_xR^2_x+p_{5}(R^2_x)^2]+\dots,
$$
$$
Q=\epsilon [q_1R^1_{xx}+q_2R^2_{xx}+q_{3}(R^1_x)^2+q_{4}R^1_xR^2_x+q_{5}(R^2_x)^2]+\dots,
$$
where $p_i,  q_i$ are some functions of $R^1, R^2$ (not all identically zero).
At the order $\epsilon$, compatibility conditions of systems (\ref{T}) and (\ref{Y}) result in eighteen constraints for the functions $f_i(R^i)$ and $p_i,  q_i$. Classification results -- up to  symmetries of system \eqref{pav} and the interchange of indices 1 and 2 --  are summarised below (five different cases).

\bigskip

{\bf Case 1:} $f_1(R^1)=\frac{1}{2}(R^1)^2$ and $f_2(R^2)=\frac{1}{2}(R^2)^2$. Then
{\small
\[ u=R^1+R^2,    \quad w=(R^1)^2+R^1 R^2+(R^2)^2,  \] \vspace{-0.7cm}
\begin{align*}
R^1_t=(2 R^1+R^2) R^1_x +P, & \qquad \qquad  R^1_y=(3 (R^1)^2+2 {R^1} {R^2}+(R^2)^2) R^1_x +R^1 P+Q,\\
R^2_t= (R^1+2R^2)R^2_x-P, & \qquad \qquad R^2_y=(3 (R^2)^2+2 {R^1} {R^2}+(R^1)^2) R^2_x-R^2 P-Q,
\end{align*}}
where 
{\footnotesize \begin{align*}
P&=\frac{\epsilon}{R^1-R^2}  \left((R^1_x)^2 \alpha '+\alpha R^1_{xx}-(R^2_x)^2 \beta'- \beta  R^2_{xx} \right)+\dots, \\
Q&= \frac{\epsilon}{R^1-R^2} \left(  \left(2 \alpha +\alpha ' (2 {R^1}+{R^2})\right)(R^1_x)^2+(\alpha -\beta ) R^1_x R^2_x+\alpha  (2 {R^1}+{R^2}) R^1_{xx}- \right.\\
&\qquad \qquad\qquad\qquad\qquad\qquad 
\left. -\beta ({R^1}+2 {R^2})  R^2_{xx}-(2 \beta +\beta '({R^1}+2 {R^2}) )(R^2_x)^2  \right)+\dots ,
\end{align*}}
and $\alpha=\alpha(R^1),\ \beta=\beta(R^2)$ are arbitrary functions. 

Further calculations show that all coefficients at the order $\epsilon^2$ are uniquely expressible in terms of the functions $\alpha$ and $\beta$.

\bigskip

{\bf Case 2:} $f_1(R^1)=\frac{1}{2}(R^1)^2$ and $f_2(R^2)=-\frac{1}{2}(R^2)^2$. Then
{\small
\[u=R^1+R^2,   \quad w=(R^1)^2+R^1 R^2,  \] \vspace{-0.7cm}
\begin{alignat*}{3}
& R^1_t &&=( 2 R^1+R^2) R^1_x +P,  \qquad \qquad  && R^1_y=(3 (R^1)^2+2 {R^1} {R^2}) R^1_x +R^1 P+Q,\\
& R^2_t && = R^1 R^2_x-P,  \qquad \qquad  && R^2_y =(R^1)^2 R^2_x+R^2 P-Q,
\end{alignat*}}
where 
{\footnotesize \begin{align*}
P&=\epsilon \left(\alpha  ({R^1}+{R^2}) R^1_{xx}+  \left(2 \alpha +\alpha ' ({R^1}+{R^2})\right)(R^1_x)^2+2 \alpha  R^1_x R^2_x+\frac{ \left(\beta ' ({R^1}+{R^2})-2 \beta \right)(R^2_x)^2+\beta  ({R^1}+{R^2}) R^2_{xx}}{({R^1}+{R^2})^2} \right)+\dots, \\
Q&= \frac{\epsilon}{(R^1-R^2)^2} \left(\alpha  (2 R^1+R^2) (R^1+R^2)^3 R^1_{xx}+\beta R^1 (R^1+R^2) R^2_{xx}+ 
({R^1}+{R^2})^2 \left( \alpha (6  {R^1}+4 R^2)+ \right. \right.\\
&\left. \left.+\alpha ' (R^1+R^2) (2 R^1+R^2) \right)(R^1_x)^2 +(R^1+R^2) \left(\beta +\alpha  (R^1+R^2) (3 R^1+R^2)\right) R^1_x R^2_x+\left(\beta ' (R^1+R^2)-2 \beta \right) R^1 (R^2_x)^2  )\right)+\dots,
\end{align*}}
and $\alpha=\alpha(R^1),\ \beta=\beta(R^2)$ are arbitrary functions. Note that the  second characteristic speed in the $t$-flow is linearly degenerate. 

Further calculations show that conditions at the order $\epsilon^2$ imply $\beta=0$, furthermore, two new arbitrary functions appear at that order.

\medskip



\bigskip

{\bf Case 3:} $f_1(R^1)=-\frac{1}{2}(R^1)^2$ and $f_2(R^2)=-\frac{1}{2}(R^2)^2$. Then
{\small
\[ u=R^1+R^2,    \quad w=R^1 R^2, \] \vspace{-0.7cm}
\begin{align*}
R^1_t=R^2 R^1_x +P, & \qquad \qquad  R^1_y=\epsilon (R^1-R^2)( (R^1-R^2) \left(\alpha R^1_{xx} +\alpha ' (R^1_x)^2 \right) -(\alpha +\beta ) R^1_x R^2_x) +\dots,\\
R^2_t= R^1 R^2_x-P, & \qquad \qquad R^2_y=\epsilon (R^2-R^1)(\left(R^1-R^2\right) \left(\beta  R^2_{xx}+\beta ' (R^2_x)^2\right)+(\alpha +\beta ) R^1_x R^2_x)+\dots,
\end{align*}}
where 
{\footnotesize \begin{align*}
P&=-\epsilon  \left(\alpha ' (R^1-R^2) (R^1_x)^2-2 (\alpha +\beta ) R^1_x R^2_x-(R^1-R^2) \left(\beta ' (R^2_x)^2+\beta  R^2_{xx}\right)+\alpha  (R^1-R^2) R^1_{xx}\right) +\dots
\end{align*}}
and $\alpha=\alpha(R^1),\ \beta=\beta(R^2)$ are arbitrary functions. Note that both characteristic speeds in the $t$-flow are linearly degenerate, while both characteristic speeds in the $y$-flow are identically zero: $\mu^1=\mu^2=0$. 

Further calculations show that conditions at the order $\epsilon^2$ imply that both functions $\alpha$ and $\beta$ must be zero (so that all $\epsilon$-terms vanish), furthermore, four  new arbitrary functions appear at that order.


\bigskip

{\bf Case 4:} $f_1(R^1)$ is arbitrary and $f_2(R^2)=\frac{1}{2}(R^2)^2$. Then
{\small
\[ u=R^1+R^2,    \quad w=f_1 +\frac{1}{2}(R^1)^2+R^1 R^2+ (R^2)^2, \]  \vspace{-0.7cm}
\begin{alignat*}{3}
&R^1_t&&=(R^1+R^2+f_1') R^1_x +P,  \qquad \qquad  &&R^1_y=\left(\frac{1}{2}(R^1)^2+R^1 R^2+ (R^2)^2+f_1+f_1'(R^1+R^2+f_1')\right)R^1_x+f_1'P+Q,\\
& R^2_t&&= (R^1+2 R^2) R^2_x-P,  \qquad \qquad && R^2_y=\left(\frac{1}{2}(R^1)^2+2R^1 R^2+ 3(R^2)^2+f_1\right) R^2_x-R^2 P-Q,
\end{alignat*}}
where 
{\footnotesize \begin{align*}
P&=\epsilon  \left(\frac{\partial {p_2}}{\partial{R^2}} (R^2_x)^2+\frac{{p_2} \left(\left(1-{f_1}''\right) R^1_x R^2_x+R^2_{xx} \left({R^2}-{f_1}'\right)+(R^2_x)^2\right)}{R^2-f_1'} \right)+\dots, \\
Q&=\epsilon  \left( \left(\left(2 R^2+R^1\right) \left(\frac{\partial {p_2}}{\partial{R^2}}+\frac{{p_2}}{R^2-f_1'}\right)+2 p_2\right)(R^2_x)^2+p_2  \left(\frac{\left(2 R^2+R^1\right) \left(f_1''-1\right)}{f_1'-R^2}+f_1''\right)R^1_x R^2_x+p_2 \left(2 R^2+R^1\right) R^2_{xx} \right)+\dots,
\end{align*}}
and $p_2(R^1, R^2)$ solves the PDE 
$$(f_1'-R^2)\frac{\partial {p_2}}{\partial{R^1}}+p_2=0.$$

\medskip

\noindent Note that if $f_1(R^1)=\frac{1}{2}(R^1)^2$ then $p_2=\frac{\beta(R^2)}{R^2-R1}$ and we recover
Case 1 with $\alpha(R^1)=0$. 

Further calculations show that conditions at the order $\epsilon^2$ imply that  either (a) $p_2\ne 0$, in which case all coefficients at $\epsilon^2$ can be uniquely expressed in terms of $f_1$ and $p_2$, or (b) $p_2=0$  (this implies that all $\epsilon$-terms vanish) and $f_1''=2$, in which case further arbitrary functions appear at that order.


\bigskip

{\bf Case 5:} $f_1(R^1)$ is arbitrary and $f_2(R^2)=-\frac{1}{2}(R^2)^2$. Then
{\small
\[
u=R^1+R^2,    \quad w=f_1 +\frac{1}{2}(R^1)^2+R^1 R^2, \] \vspace{-0.7cm}
\begin{alignat*}{3}
&R^1_t&&=(R^1+R^2+f_1') R^1_x +P,  \qquad \qquad  &&R^1_y=\left(\frac{1}{2}(R^1)^2+R^1 R^2+f_1+f_1'(R^1+R^2+f_1')\right)R^1_x+f_1'P+Q,\\
& R^2_t&&= R^1 R^2_x-P,  \qquad \qquad &&R^2_y=\left(\frac{1}{2}(R^1)^2+f_1\right) R^2_x+R^2 P-Q,
\end{alignat*}}
where 
{\footnotesize \begin{align*}
P&=\epsilon  \left(\frac{\partial {p_2}}{\partial{R^2}} (R^2_x)^2+\frac{p_2 \left(R^2_{xx} \left(f_1'+R^2\right)+\left(f_1''-1\right) R^1_x R^2_x-(R^2_x)^2\right)}{f_1'+R^2} \right)+\dots, \\
Q&=\frac{\epsilon}{f_1'+R^2}  \left( {R^1  \left(\frac{\partial {p_2}}{\partial{R^2}} \left(f_1'+R^2\right)-p_2\right)(R^2_x)^2+p_2  \left(f_1'' \left(f_1'+R^2+R^1\right)-R^1\right)R^1_x R^2_x+p_2 R^1  \left(f_1'+R^2\right)R^2_{xx}} \right)+\dots,
\end{align*}}
and $p_2(R^1, R^2)$ solves the PDE 
$$(f_1'+R^2)\frac{\partial {p_2}}{\partial{R^1}}+p_2=0.$$

\medskip

\noindent Note that if $f_1(R^1)=\frac{1}{2}(R^1)^2$ then $p_2=\frac{\beta(R^2)}{R^1+R^2}$ and we recover  Case 2 with $\alpha(R^1)=0$. Here again the  second characteristic speed in the $t$-flow is linearly degenerate. 

Further calculations show that conditions at the order $\epsilon^2$ imply  $p_2= 0$ (this implies  that all $\epsilon$-terms vanish), however, more arbitrary functions appear at that order. 


\bigskip

\noindent {\bf Terminating deformations: } There is only one case of  deformations terminating at the order $\epsilon$ (where all terms at $\epsilon^k, \ k\geq 2$  are identically zero).
 It is coming from Case 1 with $\alpha(R^1)+\beta(R^2)=0$. Hence, $\alpha(R^1)=-\beta(R^2)=const$, where the constant can be absorbed in $\epsilon$. This gives
{\small
\[  u=R^1+R^2, \quad w=(R^1)^2+R^1 R^2+(R^2)^2, \] \vspace{-0.5cm}
 \begin{align*}
R^1_t&=(2 R^1+R^2) R^1_x +\frac{\epsilon}{R^1-R^2}  ( R^1_{xx}+   R^2_{xx} ), \\
R^2_t&= (R^1+2R^2)R^2_x+\frac{\epsilon}{R^2-R^1}  ( R^1_{xx}+  R^2_{xx} ), \\
R^1_y&=\left(3 (R^1)^2+2 {R^1} {R^2}+(R^2)^2\right) R^1_x+\frac{\epsilon}{R^1-R^2}  (2(R^1_x)^2+2R^1_x R^2_x+2(R^2_x)^2+(3 R^1+R^2) R^1_{xx}+2 (R^1+R^2) R^2_{xx}),\\
R^2_y&=\left(3 (R^2)^2+2 {R^1} {R^2}+(R^1)^2\right) R^2_x+\frac{\epsilon}{R^2-R^1}(2(R^1_x)^2+2R^1_x R^2_x+2(R^2_x)^2+\left(3 R^2+R^1\right) R^2_{xx}+2(R^1+R^2) R^1_{xx}).
\end{align*}}
Introducing  the potential $q$ such that $u=q_x, \ w=q_t$, one can rewrite the above $t$-flow  as a  linearisable equation
$$
q_{tt}=q_tq_{xx}+3q_xq_{xt}-3q_x^2q_{xx}+\epsilon q_{xxx},
$$
which has appeared in \cite{MAS2, HSS}. The corresponding $y$-flow reduces to its  symmetry,
$$
q_y=2q_xq_t-q_x^3+\epsilon q_{xx}.
$$

\subsection{Two-component dispersive reductions}
\label{sec:disper2}

Here we assume that the expansions start with non-trivial $\epsilon^2$-terms,
{\small {$$
P=\epsilon^2 [p_1R^1_{xxx}+p_2 R^1_x R^1_{xx}+  p_{3}(R^1_x)^3  +p_4 R^2_{xxx}+p_5 R^2_x R^2_{xx}+  p_{6}(R^2_x)^3 +p_7 R^1_x R^2_{xx}  +p_8  R^2_x R^1_{xx}+p_9 R^1_x (R^2_x)^2 +p_{10}R^2_x (R^1_x)^2]+\dots,
$$
$$
Q=\epsilon^2 [q_1R^1_{xxx}+q_2 R^1_x R^1_{xx}+  q_{3}(R^1_x)^3  +q_4 R^2_{xxx}+q_5 R^2_x R^2_{xx}+  q_{6}(R^2_x)^3 +q_7 R^1_x R^2_{xx}  +q_8  R^2_x R^1_{xx}+q_9 R^1_x (R^2_x)^2 +q_{10}R^2_x (R^1_x)^2]+\dots,
$$}}
where $p_i, q_i$ are some functions of $R^1, R^2$ (not all identically zero).
At the order $\epsilon^2$, compatibility conditions of systems (\ref{T}) and (\ref{Y}) result in constraints for the functions $f_i(R^i)$ and $p_i, q_i$. Classification results -- up to  symmetries of system \eqref{pav} and the interchange of indices 1 and 2 --  result in five different cases. These cases are listed below, omitting the explicit presentation of $R^i_t, R^i_y$ due to the length of the expressions. 

\bigskip

\noindent$ $ {\bf  Case 1:} $f_1(R^1)= (R^1)^2$ and $f_2(R^2)=(R^2)^2$. Then
{\small
\begin{align*}
u=R^1+R^2,  &  \quad w=\frac{1}{2} (R^1+R^2)^2+(R^1)^2+(R^2)^2.  
\end{align*}}
{\bf Case 2:} $f_1(R^1)=(R^1)^2$ and $f_2(R^2)=-\frac{1}{2}(R^2)^2$. Then
{\small
\begin{align*}
u=R^1+R^2,  &  \quad w=\frac{3}{2} (R^1)^2+R^1R^2.
\end{align*}}
{\bf Case 3:} $f_1(R^1)=-\frac{1}{2}(R^1)^2$ and $f_2(R^2)=-\frac{1}{2}(R^2)^2$. Then
{\small
\begin{align*}
u=R^1+R^2,  &  \quad w=R^1R^2.
\end{align*}}
{\bf Case 4:} $f_1(R^1)$ arbitrary and $f_2(R^2)=(R^2)^2$. Then
{\small
\begin{align*}
u=R^1+R^2,  &  \quad w=\frac{1}{2} (R^1+R^2)^2+(R^2)^2+f_1. 
\end{align*}}
{\bf Case 5:} $f_1(R^1)$ arbitrary and $f_2(R^2)=-\frac{1}{2}(R^2)^2$. Then
{\small
\begin{align*}
u=R^1+R^2,  &  \quad w=\frac{1}{2} R^1(R^1+2R^2)+f_1.  
\end{align*}}

\noindent {\bf Terminating deformations: } There is only one case of deformations terminating at the order  $\epsilon^2$ (where all terms at $\epsilon^k, \ k\geq 3$  are identically zero). It is coming from Case 1 with $\alpha(R^1)+\beta(R^2)=0$. Hence, $\alpha(R^1)=-\beta(R^2)=const$, where the constant can be absorbed in $\epsilon$. This gives

{\small
\[ u=R^1+R^2, \quad w=\frac{1}{2} (R^1+R^2)^2+(R^1)^2+(R^2)^2,  \] \vspace{-0.5cm}
\begin{align*}
R^1_t&=(3 R^1+R^2) R^1_x +\frac{\epsilon^2}{R^1-R^2}  ( R^1_{xxx}+   R^2_{xxx} ), \\
R^2_t&=(R^1+3R^2)R^2_x-\frac{\epsilon^2}{R^1-R^2}  ( R^1_{xxx}+   R^2_{xxx}  ), \\
R^1_y&=\frac{3}{2} \left(5 (R^1)^2+2 R^1 R^2+(R^2)^2\right) R^1_x \\
&~~~~~~~~~~~~~~~~+\frac{\epsilon^2}{R^1-R^2}  \left((5 R^1+R^2) R^1_{xxx}+3 R^1_x \left(3 R^1_{xx}+R^2_{xx}\right)+3 R^2_x \left(R^1_{xx}+3 R^2_{xx}\right)+3 (R^1+R^2) R^2_{xxx}  \right),\\
R^2_y&=\frac{3}{2} \left((R^1)^2+2 R^1 R^2+5(R^2)^2\right) R^2_x\\
&~~~~~~~~~~~~~~~~-\frac{\epsilon^2}{R^1-R^2}  \left(3 (R^1+R^2)R^1_{xxx}+3 R^1_x \left(3 R^1_{xx}+R^2_{xx}\right)+3 R^2_x \left(R^1_{xx}+3 R^2_{xx}\right)+(R^1+5R^2) R^2_{xxx}  \right).
\end{align*}}
\noindent Here the $t$-flow is equivalent to the Kaup-Boussinesq system, see \cite{Pavlov1}, eq. (30). Introducing  the potential $q$ such that $u=q_x, \ w=q_t$, one can rewrite the above $t$-flow  in the form
$$
q_{tt}=2q_tq_{xx}+4q_xq_{xt}-6q_x^2q_{xx}+2\epsilon^2 q_{xxxx}.
$$
 The corresponding $y$-flow reduces to its symmetry,
$$
q_y=3q_xq_t-2q_x^3+2\epsilon^2 q_{xxx}.
$$

\section{Higher-order reductions and linear degeneracy}
\label{sec:dKP}

The goal of this section is to justify a conjecture that  the existence of higher-order reductions of type (\ref{w}) is a specific property of {\it linearly degenerate} dispersionless integrable systems in 3D.
In Section \ref{sec:dK} we prove that, unlike Mikhalev system (\ref{pav}), the dispersionless KP (dKP) system,
\begin{equation}
u_t=uu_x+w_y, ~~~ u_y=w_x,
\label{dKP}
\end{equation}
which is not linearly degenerate, does not possess non-trivial higher-order reductions of type (\ref{w}).
To make the paper  self-contained, we recall the definition of linear degeneracy in 3D (Section \ref{sec:ld}) and discuss an example which supports the above conjecture (Section \ref{sec:Ex}).

\subsection{Non-existence of higher-order reductions of the dKP system}
\label{sec:dK}

\begin{theorem} \label{T2} dKP system (\ref{dKP})  possesses no nontrivial reductions of type  (\ref{w}). 

\end{theorem}

\centerline{\bf Proof:}

\medskip

Direct calculations show that terms at $\epsilon, \epsilon^2, \epsilon^3$ must vanish identically. Suppose that the first nonzero term occurs at $\epsilon^n$,
\begin{equation}
w=f(u)+\epsilon^nw_n+...,
\label{red_n}
\end{equation}
where $w_n$ is a homogeneous differential polynomial of degree $n$ in the $x$-derivatives of $u$. Substituting ansatz (\ref{red_n}) into (\ref{dKP}) gives
$$
u_y=f'u_x+\epsilon^n(w_n)_x+\dots, \qquad u_t=(f'^2+u)u_x+\epsilon^n\left(f'(w_n)_x+(w_n)_y\right)+\dots.
$$
Calculating the compatibility condition,  $u_{yt}=u_{ty}$, at $\epsilon^n$ one obtains that $w_n$ must belong to the kernel of the second-order operator
$$
L=XT-Y^2-uX^2-(1+f'f'')u_xX+f''u_xY,
$$
that is, $L w_n=0$. Here $X, Y, T$ denote three commuting operators of total differentiation  associated with commuting systems $u_y=f'u_x$ and $u_t=(f'^2+u)u_x$, explicitly,
$$
\begin{array}{c}
X=u_x\frac{\partial}{\partial u}+u_{xx}\frac{\partial}{\partial u_x}+\dots, \\
\ \\
Y=f'u_x\frac{\partial}{\partial u}+(f'u_{x})_x\frac{\partial}{\partial u_x}+\dots, \\
\ \\
T=(f'^2+u)u_x\frac{\partial}{\partial u}+((f'^2+u)u_x)_x\frac{\partial}{\partial u_x}+\dots. 
\end{array}
$$ 
Introducing the operators $\tilde Y=Y-f'X$ and $\tilde T=T-2f'Y+(f'^2-u)X$ one can rewrite the operator $L$ in the form
$$
L=(\tilde T-u_x)X-\tilde Y^2+f''u_x\tilde Y.
$$
Our goal is to show that, for every $f$,  the kernel of $L$ is trivial: $L w_n=0 \implies w_n=0$  (note that $L$ is independent of $n$). We adopt the standard notation for the jet variables: $(u, u_x, u_{xx}, \dots)=(u, u_1, u_2, \dots)$. We will need the following properties of the operators $\tilde Y$ and $\tilde T$:
$$
\begin{array}{c}
\tilde Y=\sum_{s=1}^{\infty} a_s\frac{\partial}{\partial u_s},\\
\ \\
\tilde T=u_1\left(u_1\frac{\partial}{\partial u_1}+3u_2\frac{\partial}{\partial u_2}+4u_3\frac{\partial}{\partial u_3}+\dots \right)+\sum_{s=1}^{\infty} b_s\frac{\partial}{\partial u_s},
\end{array}
$$
where each coefficient $a_s$ depends on the jet variables $u, u_1, \dots, u_s$, while each coefficient $b_s$ depends on the jet variables $u, u_1, \dots, u_{s-1}$ only. 

Suppose, by contradiction, that $w_n\ne 0$, let $u_k$ be the highest jet variable occurring in $w_n$ (thus, $\frac{\partial w_n}{\partial u_k}\ne 0$). Then $u_{k+1}$ will be the highest jet variable occurring in $L w_n$, indeed, $X w_n$ will contain the term $\frac{\partial w_n}{\partial u_k}u_{k+1}$, while the expression $(-\tilde Y^2+f''u_x\tilde Y) w_n $ will not contain $u_{k+1}$  due to the above property of the operator $\tilde Y$. Thus, the expression $L w_n$ will depend on $u_{k+1}$ linearly, with the coefficient at $u_{k+1}$ equal to
\begin{equation}\label{TT}
(\tilde T-u_1)\left(\frac{\partial w_n}{\partial u_k}\right)+(k+2)u_1\frac{\partial w_n}{\partial u_k}=\tilde T\left(\frac{\partial w_n}{\partial u_k}\right)+(k+1)u_1\frac{\partial w_n}{\partial u_k}.
\end{equation}
 Let $\sigma$ be the highest-order  differential monomial in $\frac{\partial w_n}{\partial u_k}$, in the standard lexicographic order. Then, due to the above form of $\tilde T$, the highest-order monomial in the expression (\ref{TT}) will have the form $cu_1\sigma$ with some positive constant $c$, and therefore $L w_n$ does not vanish. This contradiction finishes the proof. 
 
 $\hfill \square$

\subsection{Linearly degenerate systems in 3D}
\label{sec:ld}

Let us begin with a (1+1)-dimensional  quasilinear system,
\begin{equation}\label{1+1}
{\bf v}_{t}+A({\bf v}){\bf v}_{x}=0,
\end{equation}
where ${\bf v}=(v^1, ..., v^n)$ is a vector of dependent variables and  $A$ is an $n\times n$ matrix. Recall that  system (\ref{1+1}), equivalently, matrix $A$, is said to be {\it linearly degenerate} if the eigenvalues of $A$, assumed real and distinct, are constant in the direction of the corresponding  eigenvectors. Explicitly, $L_{r^i}\lambda^i=0$, no summation, where $L_{r^i}$ is the  Lie derivative of the eigenvalue $\lambda^i$ in the direction of the corresponding eigenvector $r^i$. Linearly degenerate systems are quite exceptional from the point of view of solvability of the initial value problem and have been thoroughly investigated in  literature, see e.g. \cite{R1, R2, Liu, Serre}. There exists an invariant criterion of linear degeneracy that does not appeal to eigenvalues/eigenvectors. 
Let us introduce the characteristic polynomial of $A$,
 $$
 det(\lambda E-A({\bf v}))={\lambda  }^n  +  f_1({\bf v}){\lambda
}^{n-1} +f_2({\bf v}){\lambda}^{n-2}+ \ldots + f_n({\bf v}).
$$
The condition of linear degeneracy can be represented in the form \cite{Fer},
$$
\nabla f_1~A^{n-1}+\nabla  f_2~A^{n-2}+\ldots  +\nabla  f_n=0,
$$
where $\nabla$ is the  gradient,  $\nabla f=({{\partial
f}\over {\partial v^1}},\ldots , {{\partial f}\over {\partial
v^n}})$, and $A^k$ denotes  $k$-th power of the matrix $A$. In the $2\times 2$ case this condition simplifies to
\begin{equation*}
\nabla (tr A)\, A-\nabla (det A)=0.
\label{ldeg}
\end{equation*}
Given a (2+1)-dimensional quasilinear system,
\begin{equation}\label{2+1}
{\bf v}_{t}+A({\bf v}){\bf v}_{x}+B({\bf v}){\bf v}_{y}=0,
\end{equation}
 and imposing a travelling wave reduction, ${\bf v}(x, y, t)={\bf v}(r, s)={\bf v}(x+\alpha t,\, y+\beta t)$, we obtain $(A+\lambda I){\bf v}_s+(B+\mu I){\bf v}_r=0$, which is a (1+1)-dimensional quasilinear system of type (\ref{1+1}),
\begin{equation}\label{3ld}
{\bf v}_{s}+(A+\lambda I)^{-1}(B+\mu I){\bf v}_{r}=0.
\end{equation}
We say that (2+1)-dimensional system (\ref{2+1}) is linearly degenerate if all its travelling wave reductions (\ref{3ld}) are linearly degenerate in the above sense, for arbitrary constants $\lambda, \, \mu$.
 We refer to \cite{Majda, FKK, FM} for further analytic and geometric aspects  of linear degeneracy in higher dimensions.

\subsection{Reductions of a class of dispersionless integrable systems in 3D }
\label{sec:Ex}

Let us consider  systems of the form
\begin{equation}\label{eta}
u_t=w_x, \quad u_y=\eta w_t+\psi w_x+\varphi u_x,
\end{equation}
where $\eta, \psi, \varphi$ are some functions of $u$ and $w$. In particular, the choice $\eta=1,\, \psi=-u,\, \varphi=w$ leads to the Mikhalev system, while the choice $\eta=1,\, \psi=0,\, \varphi=u$ results in the dKP system (upon relabelling $y\leftrightarrow t$). Systems (\ref{eta}) have appeared in \cite{FMN} as dispersionless limits of integrable soliton equations in 2+1 dimensions. Our goal is to justify a conjecture that the requirement of existence of nontrivial reductions of type (\ref{w}) forces system (\ref{eta}) to be linearly degenerate. In this section, we will only focus on non-triviality of the $\epsilon$-term in (\ref {w}).
Rewriting (\ref{eta})  in matrix form (\ref{2+1}),
$$
 \left(\begin{array}{c}
u  \\
w
\end{array}
\right)_x+
\left(\begin{array}{cc}
-1/\varphi & 0  \\
0 & 0  \\
\end{array}
\right) \left(\begin{array}{c}
u  \\
w
\end{array}
\right)_y+
\left(\begin{array}{cc}
\psi/\varphi & \eta/\varphi  \\
-1 & 0  \\
\end{array}
\right) \left(\begin{array}{c}
u  \\
w
\end{array}
\right)_t=0,
$$
and applying the procedure described in Section \ref{sec:ld}, one  obtains conditions of linear degeneracy:
\begin{equation}
\eta_w=0, ~~~ \psi_w+\eta_u=0, ~~~ \varphi_w+\psi_u=0, ~~~
\varphi_u=0. \label{lindeg}
\end{equation}
Conditions of dispersionless integrability of system (\ref{eta}) (we refer to \cite{FK} for a discussion of dispersionless integrability in 3D)   reduce to a system of second-order
partial differential equations for the coefficients $\varphi,
\psi$ and $\eta$  \cite{FMN}:
{\small 
\begin{align}
\varphi _{uu}&= \frac{2 \psi _w \varphi _u-\varphi _w^2-\psi _u \varphi _w}{\eta}, \quad  &\psi _{uu}&= \frac{2 \eta _w \varphi _u-\varphi _w \psi _w+\psi _u \psi _w-2 \varphi _w \eta _u}{\eta },  \quad  &\eta _{uu}&= \frac{\eta _w \left(\psi _u-\varphi _w\right)}{\eta}, \quad \nonumber \\ 
\varphi _{uw}&= \frac{\eta _w \varphi _u}{\eta},  \quad  &\psi _{uw}&= \frac{\eta _w \psi _u}{\eta }, \quad  &\eta _{uw}&= \frac{\eta _w \eta _u}{\eta }, \quad \label{FMN} \\ 
\varphi _{ww}&= \frac{\eta _w \varphi _w}{\eta }, \quad &\psi _{ww}&= \frac{\eta _w \psi _w}{\eta }, \quad  &\eta _{ww}&=\frac{\eta _w^2}{\eta }. \quad \nonumber \end{align}}

{\noindent {\bf Remark 1.} Conditions of linear degeneracy (\ref{lindeg}) can be explicitly solved in the form
$$
\eta=cu^2+\alpha u+\beta, \quad \varphi=cw^2+\gamma w+\delta, \quad \psi=-2cuw-\alpha w-\gamma u-\mu,
$$
where $c, \alpha, \beta, \gamma, \delta, \mu$ are arbitrary constants. The requirement of dispersionless  integrability imposes the additional constraint $c=0$.

\medskip

{\noindent {\bf Remark 2.} Conditions of dispersionless integrability \eqref{FMN} can also be explicitly solved  leading, up to elementary normalisations, to the seven cases presented in Table 3 below \cite{FMN}.
\begin{center}
\centerline{\footnotesize{Table 3: Integrable cases of system (\ref{eta})}}
\begin{tabular}{ | l | l | l | p{4.5cm} |} \hline
 Case & $\eta$ & $\psi$ & $\varphi$    \\ \hline 
1 & $1$ & $\gamma u$ & $\beta w-\frac{1}{2}\beta (\beta+ \gamma)u^2+\delta u$  \\ \hline
2 & $1$ & $\alpha w+\gamma e^{\alpha u}$ & $\delta e^{2\alpha u}$, \ $\alpha \ne 0$  \\ \hline
3 & $u$ & $\alpha w+\gamma u^{\alpha +1}$ & $\delta u^{2\alpha +1}$, \ $\alpha \notin \{0, -1, -1/2\}$  \\ \hline
4 & $u$ & $-2\beta u\ln u -\beta   u$ & $\beta w+\beta^2u\ln^2 u+\delta u$ \\ \hline
5 & $u$ & $-w+\gamma \ln u$ & $\delta/u$  \\ \hline
6 & $u$ & $-\frac{1}{2}w+\gamma \sqrt u$ & $\delta \ln u$  \\ \hline
7 & $e^wr(u)$ & $e^wp(u)$ & $e^wq(u)$   \\ \hline
\end{tabular}
\end{center}
In the last case 7,  the functions $p$, $q$ and $r$ satisfy the system of ODEs
$$
r''=p'-q, ~~~~ q''r=2pq'-qp'-q^2, ~~~~ p''r=2rq'-2qr'+pp'-pq.
$$
Direct substitution of reduction (\ref{w}), where we assume $f''\ne 0$,  into (\ref{eta}) and requiring commutativity of the resulting flows leads to  $a(u) =0$ (triviality of $\epsilon$-terms) in all cases apart from

\noindent (a) Case 1 with $\beta+\gamma=\delta=0$;

\noindent (b) Case 5 with $\gamma=\delta=0$.

\noindent Note that  these cases correspond to linearly degenerate systems (\ref{eta}) (one has to use the compatibility conditions at both $\epsilon$ and $\epsilon^2$ to reach this conclusion), thus supporting the above conjecture.

\section{Concluding remarks} 

The remaining challenges are as follows:

\smallskip

\noindent (a) find a general formula for the eigenvalues $\lambda$;

\smallskip

\noindent (b) for a given eigenvalue $\lambda$, prove the existence of higher-order deformations at all orders of the deformation parameter $\epsilon$, and investigate their functional freedom. Our computations suggest that only one arbitrary function enters the deformation formulae for $\lambda=1$ and $\lambda=\frac{3}{2}$, whereas for (some) negative eigenvalues $\lambda$, several arbitrary functions (possibly, infinitely many of them) may occur in expansions (\ref{w}) at higher orders of $\epsilon$.


\section*{Acknowledgements} We thank M. Pavlov for numerous discussions. We also thank the referees for useful comments.

\bigskip
\bigskip

\noindent{\bf \large Declarations}
\medskip

\noindent{\bf Conflict of interest } There is neither conflict of interest nor additional data available for this article.

\medskip
\noindent {\bf Data availability} This article appeared in arXiv:2310.20528. 

\end{document}